\documentclass[
reprint,
showkeys,
superscriptaddress,
amsmath,amssymb,
aps,
pra,
]{revtex4-2}

\usepackage[version=3]{mhchem} % Formula subscripts using \ce{}

\usepackage{amsmath}
\usepackage{amssymb}
\usepackage{graphicx}
\usepackage{flushend} % Needed for widetext
\usepackage{xcolor}
\usepackage{nicefrac}
\usepackage{dcolumn}
\usepackage{mathtools}

\DeclarePairedDelimiter\ket{\lvert}{\rangle}
\DeclarePairedDelimiterX\braket[2]{\langle}{\rangle}{#1 \delimsize\vert #2}

\usepackage[colorlinks=true, allcolors=black]{hyperref}
\usepackage{siunitx}
\usepackage{amsfonts}
\usepackage{bm}
\usepackage[T1]{fontenc}

\newcommand{\tudo}{Department of Physics, Condensed Matter Theory, TU Dortmund University, 44221 Dortmund, Germany}

\begin{document}

\title{Towards Stirling cooler operable single photon sources based on low-noise GaAs quantum dots}

\author{Maximilian Aigner}
\thanks{These authors contributed equally to this work.}
\affiliation{Institute of Semiconductor and Solid State Physics, Johannes Kepler University, 4040 Linz, Austria}
\author{Jana Schlücking}
\thanks{These authors contributed equally to this work.}
\affiliation{\tudo}
\author{Eva Schöll}
\affiliation{Institute of Semiconductor and Solid State Physics, Johannes Kepler University, 4040 Linz, Austria}
\author{Christian Weidinger}
\affiliation{Institute of Semiconductor and Solid State Physics, Johannes Kepler University, 4040 Linz, Austria}
\author{Gabriel Undeutsch}
\affiliation{Institute of Semiconductor and Solid State Physics, Johannes Kepler University, 4040 Linz, Austria}
\author{Ievgen Brytavskyi}
\affiliation{Institute of Semiconductor and Solid State Physics, Johannes Kepler University, 4040 Linz, Austria}
\author{Thomas Oberleitner}
\affiliation{Institute of Semiconductor and Solid State Physics, Johannes Kepler University, 4040 Linz, Austria}
\author{Tobias Maria Krieger}
\affiliation{Institute of Semiconductor and Solid State Physics, Johannes Kepler University, 4040 Linz, Austria}
\author{Ailton Jose Garcia Junior}
\affiliation{Institute of Semiconductor and Solid State Physics, Johannes Kepler University, 4040 Linz, Austria}
\author{Melina Peter}
\affiliation{Institute of Semiconductor and Solid State Physics, Johannes Kepler University, 4040 Linz, Austria}
\author{Thomas K. Bracht}
\affiliation{\tudo}
\author{Michał Gawełczyk}
\affiliation{Institute of Theoretical Physics, Wrocław University of Science and Technology, Wrocław, Poland}
\author{Saimon Filipe Covre da Silva}
\affiliation{Instituto de Física Gleb Wataghin, Universidade Estadual de Campinas, 13083-970 Campinas, Brazil}
\author{Santanu Manna}
\affiliation{Department of Electrical Engineering, Indian Institute of Technology Delhi, 110016 New Delhi, India}
\author{Yusuf Karli}
\thanks{Current adress: Cavendish Laboratory, University of Cambridge, Cambridge, United Kingdom}
\affiliation{Institut für Experimentalphysik, Universität Innsbruck, 6020 Innsbruck, Austria}
\author{Gregor Weihs}
\affiliation{Institut für Experimentalphysik, Universität Innsbruck, 6020 Innsbruck, Austria}
\author{Doris E. Reiter}
\email{doris.reiter@tu-dortmund.de}
\affiliation{\tudo}
\author{Armando Rastelli}
\email{armando.rastelli@jku.at}
\affiliation{Institute of Semiconductor and Solid State Physics, Johannes Kepler University, 4040 Linz, Austria}

% Keywords: Please provide a minimum of three and a maximum of seven keywords, separated by commas


\begin{abstract}
For photonic quantum technology applications, sources capable of emitting photons with indistinguishability close to unity are essential. Ideally, these sources should not require demanding cooling systems. Here, we present temperature-dependent two-photon-interference measurements on photons produced by the radiative decay of the negative trion in a low-noise GaAs quantum dot, which are in quantitative agreement with theoretical calculations accounting for carrier-phonon interactions and coupling to excited states. While at at the lowest explored temperatures the emission linewidth reaches values only \SI{6(2)}{\%} above the Fourier limit and the indistinguishability $\mathcal{I}$ between subsequently emitted photons reaches \SI{0.966(6)}{}, the latter drops to \SI{0.05(4)}{} at \SI{55}{K}. We show that this loss can be explained with the coupling with energetically close excited trion states and suggest that the photon indistinguishability at elevated temperatures can be increased by employing Purcell enhancement of the emission rate or by increasing the energy separation of the excited states. Using cavity-enhanced emission, we experimentally verify the first route and demonstrate an improvement in photon indistinguishability from \SI{0.314(25)}{} to \SI{0.80(3)}{} at 32~K, which -- to our knowledge –- is the highest reported value at such temperature. 
\end{abstract}

\maketitle

\begingroup
\renewcommand{\thefootnote}{*}
\footnotetext{These authors contributed equally to this work.}
\endgroup

\section{Introduction}

Single photon sources (SPSs) are essential building blocks for photonic quantum technologies such as quantum key distribution and linear optical quantum computing \cite{Wang2025ScalableTechnologies}. An ideal SPS generates single photons on demand with high repetition rates, near-unity purity and indistinguishability.
Spontaneous parametric down-conversion in nonlinear crystals has long served as a benchmark source, providing indistinguishable photons at room temperature, but suffers from its probabilistic nature, which limits purity and brightness \cite{Schneeloch2019IntroductionDown-conversion}. Solid-state quantum emitters~\cite{Mark_Fox_Solid_state} offer the possibility of emitting single photons ``on demand'' but often require cryogenic operation because of the interaction of the electronic states with the thermally activated lattice vibrations. Color centers in diamond or silicon carbide can operate at elevated temperatures, yet typically show high indistinguishability only at cryogenic temperatures \cite{castelletto2021silicon}. Quantum emitters hosted in large-bandgap semiconductors~\cite{Holmes_Room_temperature, Eggleton_Controlled_Epitaxy, Wang_Quantum_Emitters} or two-dimensional materials such as transition metal dichalcogenides and hexagonal boron nitride allow room-temperature operation, but still face challenges associated with dephasing and spectral diffusion \cite{Esmann2024Solid-StateMaterials}.

Among the different approaches developed over the past decades, epitaxial semiconductor quantum dots (QDs) have proven to be particularly promising \cite{Senellart2017High-performanceSources, Zhou2023EpitaxialTechnologies}. They offer deterministic single photon generation by excitation of exciton or trion states with high brightness \cite{TommAPhotons, Ding2025NatureComputing} and show the lowest multi-photon emission probability \cite{Schweickert2018On-demandSource, Hanschke2018} of any solid-state SPS. Additional advancements via embedding the QDs into diode structures have led to charge state control and blinking-free emission, allowing deterministic addressing of the desired electronic state \cite{ZhaiLow-noisePhotonics, Warburton2000OpticalRing,Somaschi2016Near-optimalState}. Precise control of the charge environment has also led to Fourier-transform-limited line widths \cite{ZhaiLow-noisePhotonics,Kuhlmann2015Transform-limitedDot} and consequently near-unity indistinguishability even for remote emitters \cite{Zhai2022QuantumDots}.  

Despite their outstanding optical properties, all the aforementioned results have been achieved only at cryogenic temperatures around \SI{4}{K} using helium-based cryostats. These systems are bulky and consume power in the kilowatt range, which restricts their use outside laboratory environments. Compact Stirling coolers represent an attractive alternative \cite{Schlehahn2018, Musial_2020} as they can reach temperatures around \SI{30}{K}, require only about \SI{100}{W} of electrical power, and have a significantly smaller footprint. Understanding how increasing temperature influences the photon benchmark properties, namely the brightness, purity and indistinguishability, is therefore essential for future implementations relying on energy-efficient cooling technologies.

For this reason, we study the temperature dependence of photon indistinguishability from the negatively charged trion in a GaAs QD using two-photon interference in a Hong-Ou-Mandel (HOM) setup. We accompany our experimental data with theoretical simulations based on a microscopic QD model that explicitly includes coupling to phonons. Earlier works on In(Ga)As QDs addressed similar questions using more phenomenological descriptions of phonon interactions \cite{Thoma2016ExploringExperiments,Reigue2017ProbingDots}.

To investigate the effect of phonon interactions on the photon indistinguishability, we minimize other indistinguishability-degrading effects. We employ resonance fluorescence to suppress dephasing induced by the excitation pulse and avoid timing jitter in the population of the trion state. Furthermore, by embedding the QD in a p-i-n diode structure, we strongly reduce charge noise both on the nanosecond timescale, relevant for the HOM experiment, as well as on the seconds timescale probed by Michelson interferometry. As a consequence, changes in the measured indistinguishability are attributed to carrier-phonon interactions. By performing measurements under different spectral filtering conditions, we are able to distinguish between different phonon-mediated processes.

Through our combined experimental and theoretical approach, we identify distinct dephasing channels and quantify their temperature-dependent impact on photon coherence. The analysis highlights the role of phonon-induced diagonal coupling and phonon-mediated coupling to higher excited (hot) states as the dominant mechanisms limiting indistinguishability at elevated temperatures. Our detailed understanding of these coupling mechanisms enables us to propose strategies for mitigating their effects and achieving high photon indistinguishability even at temperatures reachable with compact Stirling coolers.

\begin{figure*}[!htbp]
\centering
\includegraphics[]{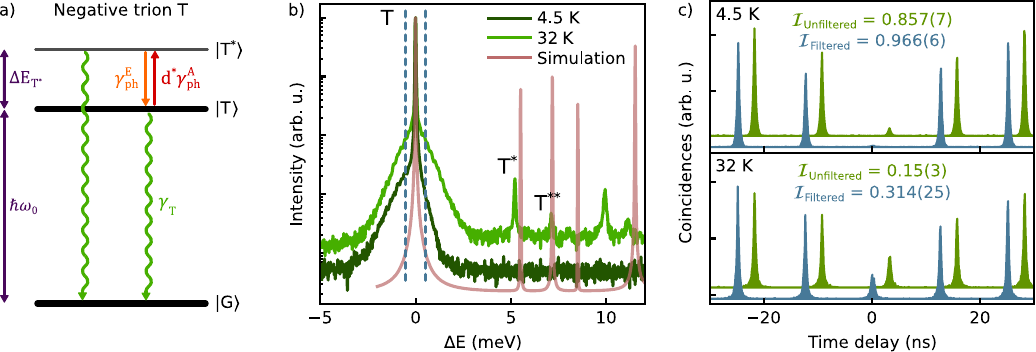}
\caption{a) Schematic few-level structure of the QD with the ground state $\ket{G}$, the negative trion state $\ket{T}$, and the first excited trion state $\ket{T^*}$. 
The green arrows mark the radiative decay channels, while the red and orange arrows indicate phonon-mediated population exchange between the trion states. 
b) Resonance fluorescence spectra of $\ket{T}$ at a temperature of \SI{4.5}{K} (dark green) and \SI{32}{K} (light green) as well as a simulated spectrum (brown). The intensity is normalized to the maximum intensity in the respective spectrum.
c) Coincidence count histograms of HOM measurements taken at \SI{4.5}{K} (top) and \SI{32}{K} (bottom) with (blue) and without (green) spectral filtering of the phonon sideband (PSB). The filter window is indicated by the blue dashed lines in b).}
\label{fig: Basics}
\end{figure*}

\section{Experimental methods}

The GaAs/AlGaAs QD used in this work was grown via droplet etching epitaxy and embedded into a p-i-n diode structure, identical to the one in Ref.~\cite{Undeutsch2025Electric-FieldDots}. The sample containing the investigated QD is mounted in a closed-cycle cryostat (``Attocube Attodry 800XS'') with a base temperature of \SI{4.5}{K}, and the temperature can be precisely adjusted via a PID-controlled heater. The temperatures reported in this manuscript correspond to readings from a sensor located below the sample. The actual sample temperatures may therefore differ slightly from the reported values. 

We apply a voltage of \SI{1}{V} to the sample, corresponding to the center of the charge plateau, where the QD contains one electron. This allows us to deterministically prepare the negative trion state $\ket{T}$, which is resonantly excited using a pulsed Ti:Sa laser (``Spectra Physics Mai Tai'') with a repetition rate of \SI{80}{MHz} and a pulse duration of about \SI{3}{ps}. The pulse area is set to $\pi$ to maximize population transfer to the trion state, and scattered laser light is suppressed using a cross-polarization scheme similar to Ref.~\cite{Kuhlmann2013AMode}.

The indistinguishability is assessed by measuring the HOM interference visibility ($\mathrm{V_{HOM}}$) of two consecutive emitted photons in an unbalanced Mach-Zehnder interferometer. In one interferometer arm, a delay of \SI{12.5}{ns} is introduced to match the laser repetition rate. Measurements were performed only in the co-polarized setting, where $V_\text{HOM}=1-2\frac{A_{0}}{\langle A_s \rangle}$, with $A_{0}$ being the coincidences of the central peak and $\langle A_s \rangle$ the averaged uncorrelated side peak area. To gain access to the indistinguishability $\mathcal{I}$, we correct $V_\text{HOM}$ for the non-zero multi-photon probability $g^{(2)}(0)$, measured in a Hanbury-Brown-Twiss setup and deviations from the 50:50 splitting ratio ($R = 0.495$, $T = 0.505$) of the interfering beam splitter \cite{Ollivier2021Hong-Ou-MandelSources}:

\begin{equation}
    \mathcal{I} \ = \ \frac{V_\text{HOM}+g^{(2)}(0)}{1-g^{(2)}(0)}\cdot\frac{R^2+T^2}{2RT}
\end{equation}

\section{Theory}
The trion in the QD is modeled as a few‑level system consisting of the ground state $|G\rangle$ (consisting of a single electron in the lowest energy state confined in the conduction band of the QD material) and the negative trion state $|T\rangle$, where both, two electrons and a hole occupy their respective lowest confined levels. In addition, we include higher‑lying excited trion configurations in which one or more carriers are in an excited state. These are denoted by $|T^*\rangle, |T^{**}\rangle$, \ldots, located at the energy $\Delta E_{T^*}$, $\Delta E_{T^{**}}$, \ldots above the ground-state trion configuration $|T\rangle$. For the numerical calculations of emission kinetics, we included only the first excited state $\ket{T^*}$, as shown in Fig.~\ref{fig: Basics}~a).

To determine the relevant energies, we directly simulate the electron, hole, and trion eigenstates of the QD using a custom implementation \cite{Gawarecki2014,Mielnik2018} of the multiband $\bm{k}\cdot\bm{p}$ method \cite{Bahder1992} combined with the configuration-interaction method for many-particle states \cite{Bryant1987} with material parameters listed in Ref.~\cite{Gawelczyk2017} and collected from Refs.~\cite{VurgaftmanJAP2001, SaidiJAP2010, TseJAP2013, AmirtharajBOOK1994}. The calculation is based on an atomic-force microscopy scan of a representative etched nanohole in AlGaAs \cite{Yuan2023}. In addition, it is possible to use the calculated eigenstates to directly simulate the carrier-phonon coupling matrix elements, including the deformation-potential and piezoelectric couplings to acoustic phonons. These can then be used to estimate the phonon relaxation rates to/from the excited states using Fermi's golden rule.

For the dynamics, the system is excited by a laser field modeled within dipole and rotating-wave approximations using a Gaussian envelope with the temporal width $\tau=3$~ps. Radiative decay of the trion is included using Lindblad operators with rate $\gamma_T$, mimicking its lifetime, we do not include emission from the excited trion state in our calculations. Photon properties such as single-photon purity and indistinguishability are obtained from two-time correlation functions following Ref.~\cite{Cosacchi2021AccuracyDot}. 

During the dynamics, the interaction of the QD and its lattice environment is given by two dominant types of phonon processes: In the absence of coupling to excited states, one mechanism is the diagonal coupling to longitudinal acoustic (LA) phonons, which leads to a loss of coherence, but also to a renormalization of the energies due to polaron formation. These non-Markovian processes are accounted for by solving the dynamics of the phonon-coupled system using process-tensor methods \cite{Cygorek2022SimulationEnvironments,  Cygorek2025UnderstandingTensors}. In the undriven system, the diagonal coupling only causes a loss of polarization and hence is often referred to as pure dephasing \cite{krummheuer2002theory}. Excellent agreement of state preparation experiments with  diagonal coupling to phonons underline the validity of this approach \cite{Kaldewey_coherent, Hanschke_reappearence, Reiter01012019}. When accounting for excited trion states, absorption and emission of phonons can promote carriers into or out of excited trion configurations; these processes are included within the Markov approximation through Lindblad dissipators. We refer to this process as \emph{cycling} through the excited states, which is known to lead to a broadening of the Zero-Phonon-Line (ZPL) \cite{Borri2001UltralongDots, Kammerer2001EfficientDots, Zibik2008IntersublevelDots}. We stress that the temperature dependence of the phonon bath also leads to a temperature-dependent ZPL broadening, as demonstrated by non-perturbative treatments \cite{Grange2009DecoherenceCalculation}. In turn, the broadening of the ZPL is directly connected to the photon indistinguishability as discussed below.

In our calculations, we assume that before excitation the system is in its electronic ground state and that the phonons are in thermal equilibrium, described by the Bose-Einstein distribution for the mean phonon number
\begin{equation}
    n(\Delta E, T)=\left(\mathrm{exp}\left[\frac{\Delta E}{k_B T}\right]-1\right)^{-1},
\end{equation}
where $k_B$ denotes the Boltzmann constant. This applies both to the diagonal phonon coupling and to the Lindblad rates associated with transitions to excited trion states. The phonon absorption rate $\gamma_{\text{ph}}^\text{A}$ and emission rate $\gamma_\text{ph}^\text{E}$ are defined through the temperature-independent phonon relaxation rate $\gamma^0_\text{ph}$ as follows:
\begin{equation}\label{eq:gamma}
    \gamma_\text{ph}^\text{A}=\gamma^0_\text{ph}n(\Delta E_{T^*},T),  \quad 
    \gamma_\text{ph}^\text{E}=\gamma^0_\text{ph}[1+n(\Delta E_{T^*},T)] \,.
\end{equation}
The degeneracy of excited trion configurations is included by multiplying the absorption rate by a factor $d^{*}$.

We use the standard material parameters for the diagonal coupling \cite{Krummheuer2005PureGeometry}. The energetic position of the excited state energy $\Delta E_{T^*}$ is taken from experiment and is in agreement with the theoretical multiband calculations, see Fig.~\ref{fig: Basics}b), which also defines the degeneracy factor. The only fitting parameters are the electron and hole confinement lengths, which are determined by fitting the low-temperature results and the phonon sidebands (PSB) in the spectra, as well as the phonon relaxation rate $\gamma_{\text{ph}}^0$. 
Details on the QD modeling, equations of motion, and the parameters used in the simulation are discussed in the supplementary information.

\begin{figure*}[htb]
\centering
\includegraphics[]{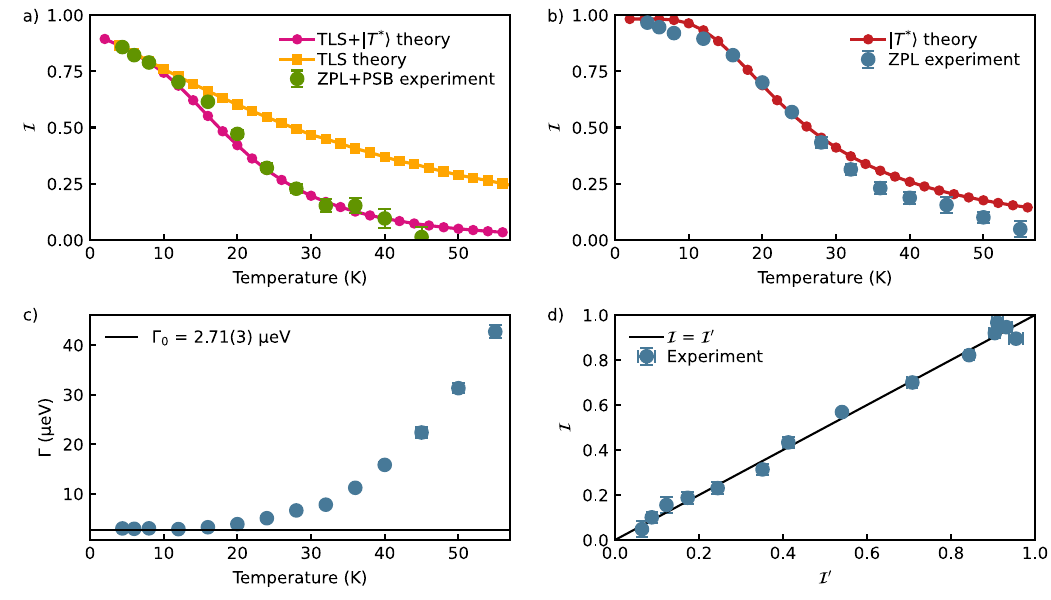}
\caption{a) Indistinguishability of the unfiltered spectrum as a function of temperature (green circles) together with a simulation including only diagonal coupling (orange squares) as well as diagonal and excited trion state coupling (pink circles) b) Indistinguishability of the filtered spectrum as a function of temperature (blue circles) and a simulation including excited trion state coupling c) Measured linewidth (blue circles) and Fourier-transform limit (black line) of the ZPL as a function of temperature d) Measured indistinguishability and $\Gamma_0/\Gamma$ of the ZPL for each temperature. The black line indicates $\mathcal{I}=\Gamma_0/\Gamma$ }
\label{fig:HOM}
\end{figure*}

\section{Results}

Before analyzing the indistinguishability, we briefly discuss the spectra of the investigated QD. Two characteristic resonance fluorescence spectra collected at \SI{4.5}{K} and \SI{32}{K} are shown in Fig.~\ref{fig: Basics}~b). 
At \SI{4.5}{K} we find the ZPL at $\Delta E=0$ superimposed on an asymmetric PSB \cite{PhysRevB.65.195313}. The ZPL contributes $\approx\SI{95}{\%}$ to the total intensity, while the PSB contains only $\approx\SI{5}{\%}$ and no additional features are visible in the spectrum at \SI{4.5}{K}.

At \SI{32}{K} the sidebands have become almost symmetric and now contribute roughly \SI{30}{\%} to the total intensity. Note that we have shifted both spectra to the same origin, thereby compensating the temperature-dependent red-shift of the ZPL at \SI{32}{K}. The respective unshifted data are shown in the supplementary material. More remarkably, besides the ZPL and PSB, four weaker emission lines appear at $\Delta E=\SI{5.1}{meV}$, $\SI{7.0}{meV}$, $\SI{9.8}{meV}$ and $\SI{11.1}{meV}$, which we attribute to excited trion states that become thermally occupied. This assumption is supported by the fact that our diode structure allows us to deterministically set the charge state of the QD and also by a simulated spectrum, which shows the same pattern of features at similar energies. This confirms that, in a single-particle picture, these features originate from configurations in which the hole occupies the first ($T^*$) or second ($T^{**}$) excited state. We stress that the simulated spectrum only shows spectral positions of the respective states, and does not reflect the line shape of the emission. 

The central focus of this work is to investigate the effect of sample temperature on the indistinguishability by measuring HOM interference between two photons subsequently emitted by a QD. Two example histograms at temperatures \SI{4.5}{K} and \SI{32}{K} are shown in Fig.~\ref{fig: Basics}~c). 
Here, we differentiate between two cases:
In the \emph{unfiltered} (green) case, light within a $\pm$ \SI{20}{meV} spectral range around the ZPL was transmitted to the HOM setup using a band-pass filter, thereby including the PSB and emission from the excited states. 
In the \emph{filtered} (blue) case, all spectral contributions except the ZPL were filtered out using narrow-band notch filters, indicated by the blue box in Fig.~\ref{fig: Basics}~b), so that mostly the ZPL interfered in the HOM setup. 
For the measurement at \SI{4.5}{K}, the suppression of the central peak at zero time delay is strong, indicating a high degree of indistinguishability for both the filtered and unfiltered emission. However, at \SI{32}{K} the coincidence counts at the central peak are increased for both measurements, indicating temperature-dependent dephasing processes and consequently a lower degree of indistinguishability.

The full temperature dependence of the indistinguishability is shown in Fig.~\ref{fig:HOM}~a), b) including both the experimental data and theoretical calculations. Data on similar measurements performed on the neutral exciton transition are shown in the supplementary information.
In Fig.~\ref{fig:HOM}~a), the resulting indistinguishability of the unfiltered emission is shown with green circles. At \SI{4.5}{K} the indistinguishability is $\mathcal{I}=0.857(7)$. With increasing temperature, it decreases monotonically, ending with $\mathcal{I}=0.000(29)$ at \SI{50}{K}.

In the theoretical modeling, we start by including only the impact of diagonal coupling in a two-level system (TLS, orange squares), neglecting the excited state for now. For temperatures $<\SI{10}{K}$, we find excellent agreement, indicating that the impact of the diagonal coupling to the LA phonons, which is responsible for the formation of the PSB, is the dominant factor in decreasing the indistinguishability at low temperatures. Note that even at temperatures very close to the absolute zero, phonon emission is always possible, such that the PSB can never be suppressed completely, limiting the achievable indistinguishability.

Including the excited state $\ket{T^*}$ in the theoretical model leads to an excellent agreement between simulations and experimental data, even at higher temperatures (pink circles). At temperatures $>\SI{10}{K}$, phonon absorption leads to a finite occupation of the excited states, resulting in an additional degradation mechanism for the indistinguishability.

In our calculations, we explicitly include the excited state $\ket{T^*}$ in our system, extending the two-level trion system to an effective three-level model. Such interactions with excited states can be treated as \textit{virtual transitions} \cite{Iles-Smith2017PhononSources}, where excited-state effects are added to the two-level system using Lindblad rates mimicking quadratic phonon processes \cite{Muljarov2004DephasingPhonons}.

The measurement results for the filtered spectrum are shown in Fig.~\ref{fig:HOM}~b). The indistinguishability also decreases with increasing temperature, following a trend similar to that observed for the unfiltered emission.
The initial value at \SI{4.5}{K} is $\mathcal{I}=0.966(6)$, showing near unity indistinguishability for subsequently emitted photons separated by \SI{12.5}{ns}. Because the PSB is filtered out, the diagonal coupling to phonons causing the PSB is neglected in the theoretical calculations, and the temperature dependence only enters via the phonon-mediated transitions to the excited trion state.

The three data points up to \SI{8}{K} show a slight linear decrease, while the theory shows almost no drop. For temperatures higher than \SI{10}{K}, the visibility decreases sharply, approaching the distinguishable regime, with $\mathcal{I}=0.048(35)$ at \SI{55}{K}. As in the unfiltered data, at elevated temperatures, transitions to the excited state tend to become more likely, thereby delaying the photon emission, leading to a loss of indistinguishability. The theoretical simulations (red circles) agree excellently with the experimental data. We stress that we chose the same parameter values regarding the excited state in both the unfiltered and filtered data.

In comparison, the resulting indistinguishability of the filtered data remains substantially higher over the entire temperature range compared to the unfiltered case, further emphasizing the role of diagonal coupling to phonons as a major source of dephasing. At \SI{4.5}{K} the indistinguishability is already $0.109(9)$ lower in the unfiltered case, confirming the benefit of spectral filtering on the resulting indistinguishability.

Additionally, there is an increasing loss of brightness with increasing temperature. First, a larger fraction of light is emitted into the PSB, leading to a lower fraction of indistinguishable photons when spectral filtering is employed \cite{PhysRevB.65.195313}. Second, the increasing number of phonons leads to a faster dampening of Rabi rotations (see supplementary information) and consequently lower probability to populate the trion state \cite{ramsay_rabi_dampening}.

The two different aspects of phonon coupling we consider here, namely the standard diagonal and the excited-state coupling, have different impacts on the spectra: While the diagonal coupling causes the appearance of the PSB, the cycling through the excited state leads to a broadening of the ZPL, both causing a loss of indistinguishability. While the PSB can be filtered, the cycling is always present. 
Accordingly, the maximum achievable indistinguishability $\mathcal{I'}$ for large time delays between the photons can be described in terms of ZPL linewidth by \cite{Bylander2003}: 
\begin{equation}\label{eq:v_hom_max}
    \mathcal{I'}=\frac{\Gamma_0}{\Gamma},\
\end{equation}
where $\Gamma_0=\frac{\hbar}{\tau}$ is the Fourier-transform-limited linewidth with $\tau$ being the lifetime of the trion state, and $\Gamma$ is the measured linewidth of the ZPL. 
To assess the broadening of the ZPL, we performed linewidth measurements via Michelson interferometry. The results are shown in Figure~\ref{fig:HOM}~c). For temperatures $<\SI{16}{K}$ the measured values are close to the Fourier-transform limit of \SI{2.71(3)}{\micro eV}. We therefore assume no significant charge noise even on the long timescales taken by the measurement (about \SI{1}{s}). This low noise environment allows to focus exclusively on the effect of temperature on the ZPL broadening. From \SI{16}{K} onwards, the linewidth increases up to \SI{42.7(1.3)}{\micro eV} at \SI{55}{K}. In theory, we find a similar but slower exponential growth of the ZPL broadening (see supplementary information). We attribute this slower growth of the theoretical data to additional effects we do not account for in the model. 

In Figure~\ref{fig:HOM}~d), the measured indistinguishability of the filtered emission is shown as a function of $\frac{\Gamma_0}{\Gamma}$, together with a straight line with slope one where $\mathcal{I}=\mathcal{I'}$, which serves as a guide to the eye. The data exhibit an approximately linear relationship, with $\mathcal{I}\approx\mathcal{I}'$ for most of the data points. With this analysis, we are able to confirm the assumption in Eq.~\eqref{eq:v_hom_max} for the investigated QD.
The impact of broadening effects on the ZPL linewidth can simply be described by:
\begin{equation}
    \Gamma=\Gamma_0+ \gamma_{\text{ZPL}},    
\end{equation}
where $\gamma_{\text{ZPL}}$ is an additional linewidth broadening induced by other dephasing processes. In Ref.~\cite{Thoma2016ExploringExperiments}, this broadening is derived by assuming the Markovian Lindblad model for the effective two-level model. We can extract $\gamma_{\text{ZPL}}$ from the experimental data and deduce from our model that $\gamma_{\text{ZPL}}$ is produced by the phonon-mediated coupling to the excited states.
Using Eq.~\ref{eq:v_hom_max}, we obtain
\begin{equation}\label{hom_vis_temp}
    \mathcal{I'}=\frac{\Gamma_0}{\Gamma}=\frac{1}{1+\gamma_{\text{ZPL}}/\Gamma_0},
\end{equation}
explicitly showing that the phonon-induced broadening is the main degrading effect on the indistinguishability. While there could also be thermally activated charge noise, this would lead to additional Gaussian line broadening. However, we find that the line shape remains Lorentzian, confirming the absence of charge noise also at higher temperatures.

From Eq.~\eqref{hom_vis_temp} it is evident that in order to increase $\mathcal{I'}$, $\gamma_{\text{ZPL}}/\Gamma_0$ is required to approach zero. Both, decreasing $\gamma_{\text{ZPL}}$ and increasing $\Gamma_0$ would lead to an increase in $\mathcal{I}$. Therefore, Purcell-enhanced emission rates are beneficial for the coherence properties and consequently indistinguishable photon generation at increased temperatures of about \SI{30}{K} \cite{Grange2017ReducingElectrodynamics, Brash2023NanocavityK}. Coupling the emitter to an optical cavity enhances the radiative recombination rate through the Purcell effect, thereby shortening the excited-state lifetime and increasing $\Gamma_0$.
\begin{figure}[htb]
\centering
\includegraphics[]{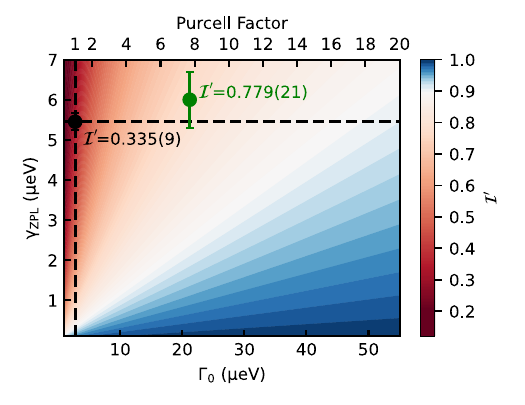}
\caption{Predicted photon indistinguishability after Eq.~\eqref{hom_vis_temp} of the PSB-filtered emission at \SI{32}{K} for different $\Gamma_0$ (bottom axis), corresponding Purcell factor (top axis), and $\gamma_{\text{ZPL}}$. The black circle indicates the data obtained from the measurements presented in Fig.~\ref{fig:HOM}, and the vertical and horizontal lines indicate a Purcell factor of one and a constant $\gamma_{\text{ZPL}}$, respectively. The green circle indicates another measurement on a different charge-tunable QD embedded in an optical microcavity~\cite{brytavskyi2026diodenanocavityfastefficient} with an estimated Purcell factor of \SI{7.7(1)}{}. The corresponding indistinguishability of $\mathcal{I}'=0.80(3)$is fully consistent with the predictions ($\mathcal{I}'=0.779(21)$).}
\label{fig:v_hom_max}
\end{figure}
We thus quantify the values of $\gamma_{\text{ZPL}}$ and $\Gamma_0$ that would lead to an indistinguishability acceptable for quantum technology applications for temperatures reachable with commercially available Stirling coolers. As a starting point, we extract $\gamma_{\text{ZPL}}$ for a temperature of \SI{32}{K} from the data in Fig.~\ref{fig:HOM}~c). In Figure~\ref{fig:v_hom_max} we plot Eq.~\eqref{eq:v_hom_max} for varying $\Gamma_0$ and $\gamma_{\text{ZPL}}$, where $\mathcal{I'}$ is color coded. We set a benchmark of $\mathcal{I'}=0.9$ shown in white. Every value above/below is shown in shades of blue/red. The black circle indicates values obtained from the  measurements presented above, corresponding to $\gamma_{ZPL}=\SI{5.5(2)}{\micro eV}$. The horizontal and vertical lines represent constant values of $\gamma_\text{ZPL}$ and $\Gamma_0$, respectively.

To test the validity of this picture experimentally, we measure $V_\text{HOM}$ at \SI{32}{K} using the trion transition from another QD embedded in a circular Bragg grating resonator cavity \cite{brytavskyi2026diodenanocavityfastefficient}. As for the experiments shown above, charge noise reduction is achieved by integrating a diode structure in the resonator, which allows us achieving nearly Fourier-transform limited emission around 4~K. The Purcell-enhanced emission from this cavity results in a reduced radiative lifetime of $\tau=\SI{31(1)}{ps}$, corresponding to $\Gamma_0=\SI{21.2(7)}{\micro eV}$.
The corresponding indistinguishability at 32~K is $\mathcal{I}=0.800(28)$ (see supplementary information), which is a substantial improvement compared to the non-cavity case and, to the best of our knowledge, the highest reported so far for a QD operated at this temperature. 
From the measurement data, we can again extract the phonon dephasing rate $\gamma_{ZPL}=\SI{6.0(7)}{\micro eV}$, which is $\SI{0.5}{\micro eV}$ higher than the other QD. We attribute the increased dephasing rate to the fact that the first excited state is located only \SI{3}{meV} above the ground state trion (to be compared with \SI{5.1}{meV} for the QD discussed above) probably resulting from a slightly different QD geometry, and consequently a larger coupling.

In order to increase the indistinguishability further, assuming a constant $\gamma_{\text{ZPL}}$ as in Fig.~\ref{fig:v_hom_max}, a Purcell factor $\geq 17$ would be required for the QDs studied here to obtain $\mathcal{I'}>0.9$. As demonstrated experimentally in Ref.\cite{rickert_high_purcell} for InGaAs QDs (with intrinsically narrower $\Gamma_0$, this type of cavity can lead to a Purcell factor $> 25$, thereby enabling $\mathcal{I'}>0.9$ at $\approx\SI{30}{K}$.
Decreasing the dephasing via $\gamma_\text{ZPL}$ is a less trivial task. 
If $\Gamma_0$ is kept constant, $\hbar\gamma_\text{ZPL}$ would need to be reduced to $\SI{0.31}{\micro eV}$ to yield the same result. This can possibly be achieved by increasing the energy splitting of the excited states, thereby decreasing the interaction strength. The energy splitting is largely dependent on the QD size, where a smaller QD leads to a larger splitting. In Ref.\cite{Denning2020PhononSources}, a decrease in the dephasing rate for QDs $> \SI{5}{nm}$ was reported under the assumption of a spherically symmetric QD with a harmonic potential, due to the complex dependency of phonon spectral density and excited state separation on the size. We stress that these approximations are quite crude, and a realistic QD model might lead to a different size-dependent behavior.

\section{Conclusion}
In conclusion, we have studied the temperature dependence of the indistinguishability of photons emitted by low-noise GaAs QDs, quantifying the impact of phonons on the process. At \SI{4.5}{K} we measure indistinguishability values as high as $\mathcal{I}=$ \SI{0.966(6)}{}. With increasing temperature, a monotonic drop is observed, which we attribute to cycling through excited states. Without further measures, the indistinguishability reaches zero at about \SI{55}{K} for the spectrally filtered, and at \SI{45}{K} for the unfiltered emission. With this in mind, we discuss possible mitigation strategies to reach high indistinguishability at temperatures around \SI{30}{K}, achievable with commercially available Stirling coolers. By using a sample featuring an estimated Purcell enhancement of the emission rate by \SI{7.7(1)}{}, we were able to increase the indistinguishability of the filtered emission at \SI{32}{K} from 0.314(25) to 0.80(3). A further increase of the Purcell factor to 17 would lead to an indistinguishability exceeding 0.9. While the drop in indistinguishability cannot be completely traced back to excited state coupling, we expect their influence can be further reduced by increasing the energy separation to the trion state. This, in turn, can be achieved by manipulating the QD geometry \cite{williamson_electronic,Navarez_electronicstructure}. With this combined experimental and theoretical effort, QD-based SPSs can be optimized for Stirling cooler operation, opening new opportunities for their integration into practical quantum technology applications relying on two-photon interference.

\medskip
\textbf{Acknowledgements} \par 
We acknowledge the expert support of M. Rota, R. Trotta, Q. Buchinger, T. Huber-Loyola, S. Höfling during the initial work leading to the microcavity sample used for the data point in Fig. 3.  
JS, TKB and DER acknowledge funding from the Deutsche Forschungsgemeinschaft (DFG, German Research Foundation) under project BRAIDS (560765700). DER acknowledges funding from the Federal Ministry of Research, Technology and Space within the QuantERA project MEEDGARD (16KIS2058).

M.\,G. acknowledges the financing of the MEEDGARD project funded within the QuantERA II Program that has received funding from the European Union's Horizon 2020 research and innovation program under Grant Agreement No. 101017733 and the National Centre for Research and Development, Poland --- project No. QUANTERAII/2/56/MEEDGARD/2024.
M.\,G. is grateful to Krzysztof Gawarecki for sharing his computational code.
Part of the calculations have been carried out using resources provided by the Wroc{\l}aw Centre for Networking and Supercomputing (M.\,G.).

MA, ES, CW, GU, MP, AG and AR acknowledge the European Union HE EIC Pathfinder challenges action under grant agreement No. 101115575, the QuantERA II program that has received funding from the European Union’s Horizon 2020 research and innovation program under Grant Agreement No. 101017733 via the projects QD-E-QKD and MEEDGARD (FFG Grants No. 891366 and 906046) the Austrian Science Fund FWF via the projects 10.55776/PIN9997324 , 10.55776/FG5, 10.55776/I4380, 10.55776/PIN4389523, 10.55776/F7113, and from the Cluster of Excellence quantA [10.55776/COE1] as well as the Linz Institute of Technology (LIT),  via the LIT project LIT-2025-14-YOU-121, and the LIT Secure and Correct Systems Lab, supported by the State of Upper Austria and the Austrian Federal Ministry of Education, Science and Research.

S.F.C. da Silva acknowledges São Paulo Research Foundation (FAPESP), Brasil, Process Number 2024/08527-2 for financial support.

\medskip

\bibliography{refs,refs_MG,refs_others}

@article{TommAPhotons,
author={Tomm, Natasha
and Javadi, Alisa
and Antoniadis, Nadia Olympia
and Najer, Daniel
and L{\"o}bl, Matthias Christian
and Korsch, Alexander Rolf
and Schott, R{\"u}diger
and Valentin, Sascha Ren{\'e}
and Wieck, Andreas Dirk
and Ludwig, Arne
and Warburton, Richard John},
title={A bright and fast source of coherent single photons},
journal={Nature Nanotechnology},
year={2021},
month={Apr},
day={01},
volume={16},
number={4},
pages={399-403},
issn={1748-3395},
doi={10.1038/s41565-020-00831-x},
url={https://doi.org/10.1038/s41565-020-00831-x}
}

@article{Kuhlmann2013AMode,
    title = {{A dark-field microscope for background-free detection of resonance fluorescence from single semiconductor quantum dots operating in a set-and-forget mode}},
    year = {2013},
    journal = {Review of Scientific Instruments},
    author = {Kuhlmann, Andreas V. and Houel, Julien and Brunner, Daniel and Ludwig, Arne and Reuter, Dirk and Wieck, Andreas D. and Warburton, Richard J.},
    number = {7},
    month = {7},
    pages = {73905},
    volume = {84},
    publisher = {AIP Publishing},
    url = {/aip/rsi/article/84/7/073905/360610/A-dark-field-microscope-for-background-free},
    doi = {10.1063/1.4813879/360610},
    issn = {00346748},
    arxivId = {1303.2055}
}

@article{Cosacchi2021AccuracyDot,
  title = {Accuracy of the Quantum Regression Theorem for Photon Emission from a Quantum Dot},
  author = {Cosacchi, M. and Seidelmann, T. and Cygorek, M. and Vagov, A. and Reiter, D. E. and Axt, V. M.},
  journal = {Phys. Rev. Lett.},
  volume = {127},
  issue = {10},
  pages = {100402},
  numpages = {8},
  year = {2021},
  month = {Aug},
  publisher = {American Physical Society},
  doi = {10.1103/PhysRevLett.127.100402},
  url = {https://link.aps.org/doi/10.1103/PhysRevLett.127.100402}
}

@article{Grange2009DecoherenceCalculation,
  title = {Decoherence in quantum dots due to real and virtual transitions: A nonperturbative calculation},
  author = {Grange, Thomas},
  journal = {Phys. Rev. B},
  volume = {80},
  issue = {24},
  pages = {245310},
  numpages = {8},
  year = {2009},
  month = {Dec},
  publisher = {American Physical Society},
  doi = {10.1103/PhysRevB.80.245310},
  url = {https://link.aps.org/doi/10.1103/PhysRevB.80.245310}
}

@article{Kammerer2001EfficientDots,
    title = {{Efficient acoustic phonon broadening in single self-assembled {InAs/GaAs} quantum dots}},
    year = {2001},
    journal = {Phys. Rev. B},
    author = {Kammerer, C. and Cassabois, G. and Voisin, C. and Delalande, C. and Roussignol, Ph and Lema{\^{i}}tre, A. and G{\'{e}}rard, J. M.},
    number = {3},
    month = {12},
    pages = {033313},
    volume = {65},
    publisher = {American Physical Society},
    url = {https://journals.aps.org/prb/abstract/10.1103/PhysRevB.65.033313},
    doi = {10.1103/PhysRevB.65.033313},
    issn = {01631829}
}

@article{Undeutsch2025Electric-FieldDots,
author={Undeutsch, Gabriel
and Aigner, Maximilian
and Garcia Jr., Ailton J.
and Reindl, Johannes
and Peter, Melina
and Mader, Simon
and Weidinger, Christian
and Covre da Silva, Saimon F.
and Manna, Santanu
and Sch{\"o}ll, Eva
and Rastelli, Armando},
title={Electric-Field Control of Photon Indistinguishability in Cascaded Decays in Quantum Dots},
journal={Nano Letters},
year={2025},
month={Apr},
day={30},
publisher={American Chemical Society},
volume={25},
number={17},
pages={7121-7127},
issn={1530-6984},
doi={10.1021/acs.nanolett.5c01354},
url={https://doi.org/10.1021/acs.nanolett.5c01354}
}

@article{Zhou2023EpitaxialTechnologies,
author = {Xiaoyan Zhou and Liang Zhai and Jin Liu},
title = {{Epitaxial quantum dots: a semiconductor launchpad for photonic quantum technologies}},
volume = {1},
journal = {Photonics Insights},
number = {2},
publisher = {SPIE},
pages = {R07},
year = {2023},
doi = {10.3788/PI.2022.R07},
URL = {https://doi.org/10.3788/PI.2022.R07}
}

@article{Thoma2016ExploringExperiments,
  title = {Exploring Dephasing of a Solid-State Quantum Emitter via Time- and Temperature-Dependent Hong-Ou-Mandel Experiments},
  author = {Thoma, A. and Schnauber, P. and Gschrey, M. and Seifried, M. and Wolters, J. and Schulze, J.-H. and Strittmatter, A. and Rodt, S. and Carmele, A. and Knorr, A. and Heindel, T. and Reitzenstein, S.},
  journal = {Phys. Rev. Lett.},
  volume = {116},
  issue = {3},
  pages = {033601},
  numpages = {5},
  year = {2016},
  month = {Jan},
  publisher = {American Physical Society},
  doi = {10.1103/PhysRevLett.116.033601},
  url = {https://link.aps.org/doi/10.1103/PhysRevLett.116.033601}
}

@article{Senellart2017High-performanceSources,
author={Senellart, Pascale
and Solomon, Glenn
and White, Andrew},
title={High-performance semiconductor quantum-dot single-photon sources},
journal={Nature Nanotechnology},
year={2017},
month={Nov},
day={01},
volume={12},
number={11},
pages={1026-1039},
issn={1748-3395},
doi={10.1038/nnano.2017.218},
url={https://doi.org/10.1038/nnano.2017.218}
}

@article{Ollivier2021Hong-Ou-MandelSources,
  title = {Hong-Ou-Mandel Interference with Imperfect Single Photon Sources},
  author = {Ollivier, H. and Thomas, S. E. and Wein, S. C. and de Buy Wenniger, I. Maillette and Coste, N. and Loredo, J. C. and Somaschi, N. and Harouri, A. and Lemaitre, A. and Sagnes, I. and Lanco, L. and Simon, C. and Anton, C. and Krebs, O. and Senellart, P.},
  journal = {Phys. Rev. Lett.},
  volume = {126},
  issue = {6},
  pages = {063602},
  numpages = {6},
  year = {2021},
  month = {Feb},
  publisher = {American Physical Society},
  doi = {10.1103/PhysRevLett.126.063602},
  url = {https://link.aps.org/doi/10.1103/PhysRevLett.126.063602}
}

@article{Zibik2008IntersublevelDots,
  title = {Intersublevel polaron dephasing in self-assembled quantum dots},
  author = {Zibik, E. A. and Grange, T. and Carpenter, B. A. and Ferreira, R. and Bastard, G. and Vinh, N. Q. and Phillips, P. J. and Steer, M. J. and Hopkinson, M. and Cockburn, J. W. and Skolnick, M. S. and Wilson, L. R.},
  journal = {Phys. Rev. B},
  volume = {77},
  issue = {4},
  pages = {041307(R)},
  numpages = {4},
  year = {2008},
  month = {Jan},
  publisher = {American Physical Society},
  doi = {10.1103/PhysRevB.77.041307},
  url = {https://link.aps.org/doi/10.1103/PhysRevB.77.041307}
}

@article{Schneeloch2019IntroductionDown-conversion,
    title = {{Introduction to the absolute brightness and number statistics in spontaneous parametric down-conversion}},
    year = {2019},
    journal = {Journal of Optics},
    author = {Schneeloch, James and Knarr, Samuel H. and Bogorin, Daniela F. and Levangie, Mackenzie L. and Tison, Christopher C. and Frank, Rebecca and Howland, Gregory A. and Fanto, Michael L. and Alsing, Paul M.},
    number = {4},
    month = {2},
    pages = {043501},
    volume = {21},
    publisher = {IOP Publishing},
    url = {https://iopscience.iop.org/article/10.1088/2040-8986/ab05a8 https://iopscience.iop.org/article/10.1088/2040-8986/ab05a8/meta},
    doi = {10.1088/2040-8986/AB05A8},
    issn = {2040-8986},
    arxivId = {1807.10885}
}

@article{ZhaiLow-noisePhotonics,
author={Zhai, Liang
and L{\"o}bl, Matthias C.
and Nguyen, Giang N.
and Ritzmann, Julian
and Javadi, Alisa
and Spinnler, Clemens
and Wieck, Andreas D.
and Ludwig, Arne
and Warburton, Richard J.},
title={Low-noise {GaAs} quantum dots for quantum photonics},
journal={Nature Communications},
year={2020},
month={Sep},
day={21},
volume={11},
number={1},
pages={4745},
issn={2041-1723},
doi={10.1038/s41467-020-18625-z},
url={https://doi.org/10.1038/s41467-020-18625-z}
}

@article{Brash2023NanocavityK,
doi = {10.1088/2633-4356/acf5c0},
url = {https://doi.org/10.1088/2633-4356/acf5c0},
year = {2023},
month = {oct},
publisher = {IOP Publishing},
volume = {3},
number = {4},
pages = {045001},
author = {Brash, A J and Iles-Smith, J},
title = {Nanocavity enhanced photon coherence of solid-state quantum emitters operating up to 30 K},
journal = {Materials for Quantum Technology}
}

@article{Ding2025NatureComputing,
author={Ding, Xing
and Guo, Yong-Peng
and Xu, Mo-Chi
and Liu, Run-Ze
and Zou, Geng-Yan
and Zhao, Jun-Yi
and Ge, Zhen-Xuan
and Zhang, Qi-Hang
and Liu, Hua-Liang
and Wang, Lin-Jun
and Chen, Ming-Cheng
and Wang, Hui
and He, Yu-Ming
and Huo, Yong-Heng
and Lu, Chao-Yang
and Pan, Jian-Wei},
title={High-efficiency single-photon source above the loss-tolerant threshold for efficient linear optical quantum computing},
journal={Nature Photonics},
year={2025},
month={Apr},
day={01},
volume={19},
number={4},
pages={387-391},
issn={1749-4893},
doi={10.1038/s41566-025-01639-8},
url={https://doi.org/10.1038/s41566-025-01639-8}
}

@article{Somaschi2016Near-optimalState,
author={Somaschi, N.
and Giesz, V.
and De Santis, L.
and Loredo, J. C.
and Almeida, M. P.
and Hornecker, G.
and Portalupi, S. L.
and Grange, T.
and Ant{\'o}n, C.
and Demory, J.
and G{\'o}mez, C.
and Sagnes, I.
and Lanzillotti-Kimura, N. D.
and Lema{\'i}tre, A.
and Auffeves, A.
and White, A. G.
and Lanco, L.
and Senellart, P.},
title={Near-optimal single-photon sources in the solid state},
journal={Nature Photonics},
year={2016},
month={May},
day={01},
volume={10},
number={5},
pages={340-345},
issn={1749-4893},
doi={10.1038/nphoton.2016.23},
url={https://doi.org/10.1038/nphoton.2016.23}
}

@article{Schweickert2018On-demandSource,
    title = {{On-demand generation of background-free single photons from a solid-state source}},
    year = {2018},
    journal = {Applied Physics Letters},
    author = {Schweickert, Lucas and J{\"{o}}ns, Klaus D. and Zeuner, Katharina D. and Covre Da Silva, Saimon Filipe and Huang, Huiying and Lettner, Thomas and Reindl, Marcus and Zichi, Julien and Trotta, Rinaldo and Rastelli, Armando and Zwiller, Val},
    number = {9},
    month = {2},
    pages = {93106},
    volume = {112},
    publisher = {American Institute of Physics Inc.},
    url = {/aip/apl/article/112/9/093106/36016/On-demand-generation-of-background-free-single},
    doi = {10.1063/1.5020038/36016},
    issn = {00036951},
    arxivId = {1712.06937}
}

@article{Warburton2000OpticalRing,
author={Warburton, R. J.
and Sch{\"a}flein, C.
and Haft, D.
and Bickel, F.
and Lorke, A.
and Karrai, K.
and Garcia, J. M.
and Schoenfeld, W.
and Petroff, P. M.},
title={Optical emission from a charge-tunable quantum ring},
journal={Nature},
year={2000},
month={Jun},
day={01},
volume={405},
number={6789},
pages={926-929},
issn={1476-4687},
doi={10.1038/35016030},
url={https://doi.org/10.1038/35016030}
}

@article{Denning2020PhononSources,
    title = {{Phonon effects in quantum dot single-photon sources}},
    year = {2020},
    journal = {Optical Materials Express},
    author = {Denning, Emil V. and Iles-Smith, Jake and Gregersen, Niels and Mork, Jesper},
    number = {1},
    month = {1},
    pages = {222},
    volume = {10},
    publisher = {Optica Publishing Group},
    doi = {10.1364/ome.380601},
    issn = {21593930}
}

@article{Iles-Smith2017PhononSources,
    title = {{Phonon scattering inhibits simultaneous near-unity efficiency and indistinguishability in semiconductor single-photon sources}},
    year = {2017},
    journal = {Nature Photonics},
    author = {Iles-Smith, Jake and McCutcheon, Dara P.S. and Nazir, Ahsan and M{\o}rk, Jesper},
    number = {8},
    month = {8},
    pages = {521--526},
    volume = {11},
    publisher = {Nature Publishing Group},
    doi = {10.1038/nphoton.2017.101},
    issn = {17494893}
}

@article{Reigue2017ProbingDots,
  title = {Probing Electron-Phonon Interaction through Two-Photon Interference in Resonantly Driven Semiconductor Quantum Dots},
  author = {Reigue, Antoine and Iles-Smith, Jake and Lux, Fabian and Monniello, L\'eonard and Bernard, Mathieu and Margaillan, Florent and Lemaitre, Aristide and Martinez, Anthony and McCutcheon, Dara P. S. and M\o{}rk, Jesper and Hostein, Richard and Voliotis, Valia},
  journal = {Phys. Rev. Lett.},
  volume = {118},
  issue = {23},
  pages = {233602},
  numpages = {6},
  year = {2017},
  month = {Jun},
  publisher = {American Physical Society},
  doi = {10.1103/PhysRevLett.118.233602},
  url = {https://link.aps.org/doi/10.1103/PhysRevLett.118.233602}
}

@article{Krummheuer2005PureGeometry,
  title = {Pure dephasing and phonon dynamics in {GaAs- and GaN-based} quantum dot structures: Interplay between material parameters and geometry},
  author = {Krummheuer, B. and Axt, V. M. and Kuhn, T. and D'Amico, I. and Rossi, F.},
  journal = {Phys. Rev. B},
  volume = {71},
  issue = {23},
  pages = {235329},
  numpages = {13},
  year = {2005},
  month = {Jun},
  publisher = {American Physical Society},
  doi = {10.1103/PhysRevB.71.235329},
  url = {https://link.aps.org/doi/10.1103/PhysRevB.71.235329}
}

@article{Zhai2022QuantumDots,
    title = {{Quantum interference of identical photons from remote {GaAs} quantum dots}},
    year = {2022},
    journal = {Nature Nanotechnology},
    author = {Zhai, Liang and Nguyen, Giang N. and Spinnler, Clemens and Ritzmann, Julian and L{\"{o}}bl, Matthias C. and Wieck, Andreas D. and Ludwig, Arne and Javadi, Alisa and Warburton, Richard J.},
    number = {8},
    month = {5},
    pages = {829--833},
    volume = {17},
    publisher = {Nature Publishing Group},
    url = {https://www.nature.com/articles/s41565-022-01131-2},
    doi = {10.1038/s41565-022-01131-2},
    issn = {1748-3395},
}

@article{Grange2017ReducingElectrodynamics,
  title = {Reducing Phonon-Induced Decoherence in Solid-State Single-Photon Sources with Cavity Quantum Electrodynamics},
  author = {Grange, T. and Somaschi, N. and Ant\'on, C. and De Santis, L. and Coppola, G. and Giesz, V. and Lema\^{\i}tre, A. and Sagnes, I. and Auff\`eves, A. and Senellart, P.},
  journal = {Phys. Rev. Lett.},
  volume = {118},
  issue = {25},
  pages = {253602},
  numpages = {6},
  year = {2017},
  month = {Jun},
  publisher = {American Physical Society},
  doi = {10.1103/PhysRevLett.118.253602},
  url = {https://link.aps.org/doi/10.1103/PhysRevLett.118.253602}
}

@article{Wang2025ScalableTechnologies,
author={Wang, Hui
and Ralph, Timothy C.
and Renema, Jelmer J.
and Lu, Chao-Yang
and Pan, Jian-Wei},
title={Scalable photonic quantum technologies},
journal={Nature Materials},
year={2025},
month={Dec},
day={01},
volume={24},
number={12},
pages={1883-1897},
issn={1476-4660},
doi={10.1038/s41563-025-02306-7},
url={https://doi.org/10.1038/s41563-025-02306-7}
}

@article{Cygorek2022SimulationEnvironments,
    title = {{Simulation of open quantum systems by automated compression of arbitrary environments}},
    year = {2022},
    journal = {Nature Physics},
    author = {Cygorek, Moritz and Cosacchi, Michael and Vagov, Alexei and Axt, Vollrath Martin and Lovett, Brendon W. and Keeling, Jonathan and Gauger, Erik M.},
    number = {6},
    month = {3},
    pages = {662--668},
    volume = {18},
    publisher = {Nature Publishing Group},
    url = {https://www.nature.com/articles/s41567-022-01544-9},
    doi = {10.1038/s41567-022-01544-9},
    issn = {1745-2481}
}

@article{Esmann2024Solid-StateMaterials,
    title = {{Solid-State Single-Photon Sources: Recent Advances for Novel Quantum Materials}},
    year = {2024},
    journal = {Advanced Functional Materials},
    author = {Esmann, Martin and Wein, Stephen C. and Ant{\'{o}}n-Solanas, Carlos},
    number = {30},
    month = {7},
    pages = {2315936},
    volume = {34},
    publisher = {John Wiley {\&} Sons, Ltd},
    url = {https://onlinelibrary.wiley.com/doi/full/10.1002/adfm.202315936 https://onlinelibrary.wiley.com/doi/abs/10.1002/adfm.202315936 https://advanced.onlinelibrary.wiley.com/doi/10.1002/adfm.202315936},
    doi = {10.1002/ADFM.202315936},
    issn = {1616-3028}
}

@article{Kuhlmann2015Transform-limitedDot,
author={Kuhlmann, Andreas V.
and Prechtel, Jonathan H.
and Houel, Julien
and Ludwig, Arne
and Reuter, Dirk
and Wieck, Andreas D.
and Warburton, Richard J.},
title={Transform-limited single photons from a single quantum dot},
journal={Nature Communications},
year={2015},
month={Sep},
day={08},
volume={6},
number={1},
pages={8204},
issn={2041-1723},
doi={10.1038/ncomms9204},
url={https://doi.org/10.1038/ncomms9204}
}

@article{Borri2001UltralongDots,
    title = {{Ultralong Dephasing Time in {InGaAs} Quantum Dots}},
    year = {2001},
    journal = {Phys. Rev. Lett.},
    author = {Borri, P. and Langbein, W. and Schneider, S. and Woggon, U. and Sellin, R. L. and Ouyang, D. and Bimberg, D.},
    number = {15},
    month = {9},
    pages = {157401},
    volume = {87},
    publisher = {American Physical Society},
    url = {https://journals.aps.org/prl/abstract/10.1103/PhysRevLett.87.157401},
    doi = {10.1103/PhysRevLett.87.157401},
    issn = {10797114}
}

@article{Cygorek2025UnderstandingTensors,
	title = {Understanding and utilizing the inner bonds of process tensors},
	pages = {024},
	author = {Cygorek, Moritz and Gauger, Erik},
	journal = {SciPost Phys.},
	volume = {18},
	year = {2025},
	publisher = {SciPost},
	doi = {10.21468/SciPostPhys.18.1.024},
	url = {https://scipost.org/10.21468/SciPostPhys.18.1.024}
}

@misc{brytavskyi2026diodenanocavityfastefficient,
      title={A diode nanocavity for fast, efficient and tunable emission of highly entangled photon pairs and Fourier-transform-limited single photons}, 
      author={Ievgen Brytavskyi and Thomas Oberleitner and Christian Weidinger and Maximilian Aigner and Gabriel Undeutsch and Tobias Steindl and Johannes Reindl and Ailton Garcia Jr. and Melina Peter and Christian Schimpf and Santanu Manna and Michele B. Rota and Quirin Buchinger and Sven Höfling and Tobias Huber-Loyola and Rinaldo Trotta and Tobias M. Krieger and Eva Schöll and Armando Rastelli},
      year={2026},
      eprint={2607.11494},
      archivePrefix={arXiv},
      primaryClass={quant-ph},
      url={https://arxiv.org/abs/2607.11494}, 
}

@article{Bylander2003,
author={Bylander, J.
and Robert-Philip, I.
and Abram, I.},
title={Interference and correlation of two independent photons},
journal={The European Physical Journal D - Atomic, Molecular, Optical and Plasma Physics},
year={2003},
month={Feb},
day={01},
volume={22},
number={2},
pages={295-301},
issn={1434-6079},
doi={10.1140/epjd/e2002-00236-6},
url={https://doi.org/10.1140/epjd/e2002-00236-6}
}

@article{Muljarov2004DephasingPhonons,
  title = {Dephasing in Quantum Dots: Quadratic Coupling to Acoustic Phonons},
  author = {Muljarov, E. A. and Zimmermann, R.},
  journal = {Phys. Rev. Lett.},
  volume = {93},
  issue = {23},
  pages = {237401},
  numpages = {4},
  year = {2004},
  month = {Nov},
  publisher = {American Physical Society},
  doi = {10.1103/PhysRevLett.93.237401},
  url = {https://link.aps.org/doi/10.1103/PhysRevLett.93.237401}
}

@article{Holmes_Room_temperature,
author = {Holmes, Mark J. and Choi, Kihyun and Kako, Satoshi and Arita, Munetaka and Arakawa, Yasuhiko},
title = {Room-Temperature Triggered Single Photon Emission from a III-Nitride Site-Controlled Nanowire Quantum Dot},
journal = {Nano Letters},
volume = {14},
number = {2},
pages = {982-986},
year = {2014},
doi = {10.1021/nl404400d},
URL = {https://doi.org/10.1021/nl404400d}
}

@article{Eggleton_Controlled_Epitaxy,
    author = {Eggleton, Katie M. and Cannon, Joseph K. and Bishop, Sam G. and Hadden, John P. and Zhao, Chunyu and Kappers, Menno J. and Oliver, Rachel A. and Bennett, Anthony J.},
    title = {Controlled epitaxy of room-temperature quantum emitters in gallium nitride},
    journal = {APL Photonics},
    volume = {11},
    number = {1},
    pages = {016103},
    year = {2026},
    month = {01},
    issn = {2378-0967},
    doi = {10.1063/5.0300338},
    url = {https://doi.org/10.1063/5.0300338},
}

@article{Wang_Quantum_Emitters,
author = {Wang, Xiao-Jie and Zhao, Shuang and Fang, Hong-Hua and Xing, Renhao and Chai, Yuan and Li, Xiao-Ze and Zhou, Yun-Ke and Zhang, Yan and Huang, Guan-Yao and Hu, Cong and Sun, Hong-Bo},
title = {Quantum Emitters with Narrow Band and High Debye–Waller Factor in Aluminum Nitride Written by Femtosecond Laser},
journal = {Nano Letters},
volume = {23},
number = {7},
pages = {2743-2749},
year = {2023},
doi = {10.1021/acs.nanolett.3c00019},
URL = {https://doi.org/10.1021/acs.nanolett.3c00019}
}

@article{Hanschke2018,
author={Hanschke, Lukas
and Fischer, Kevin A.
and Appel, Stefan
and Lukin, Daniil
and Wierzbowski, Jakob
and Sun, Shuo
and Trivedi, Rahul
and Vu{\v{c}}kovi{\'{c}}, Jelena
and Finley, Jonathan J.
and M{\"u}ller, Kai},
title={Quantum dot single-photon sources with ultra-low multi-photon probability},
journal={npj Quantum Information},
year={2018},
month={Sep},
day={14},
volume={4},
number={1},
pages={43},
issn={2056-6387},
doi={10.1038/s41534-018-0092-0},
url={https://doi.org/10.1038/s41534-018-0092-0}
}

@article{Mark_Fox_Solid_state,
author = {Fox, A. Mark},
title = {Solid-State Quantum Emitters},
journal = {Advanced Quantum Technologies},
volume = {8},
number = {2},
pages = {2300390},
doi = {https://doi.org/10.1002/qute.202300390},
url = {https://advanced.onlinelibrary.wiley.com/doi/abs/10.1002/qute.202300390},
year = {2025}
}

@article{Schlehahn2018,
author={Schlehahn, Alexander
and Fischbach, Sarah
and Schmidt, Ronny
and Kaganskiy, Arsenty
and Strittmatter, Andr{\'e}
and Rodt, Sven
and Heindel, Tobias
and Reitzenstein, Stephan},
title={A stand-alone fiber-coupled single-photon source},
journal={Scientific Reports},
year={2018},
month={Jan},
day={22},
volume={8},
number={1},
pages={1340},
issn={2045-2322},
doi={10.1038/s41598-017-19049-4},
url={https://doi.org/10.1038/s41598-017-19049-4}
}

@article{Musial_2020,
author = {Musiał, Anna and Żołnacz, Kinga and Srocka, Nicole and Kravets, Oleh and Große, Jan and Olszewski, Jacek and Poturaj, Krzysztof and Wójcik, Grzegorz and Mergo, Paweł and Dybka, Kamil and Dyrkacz, Mariusz and Dłubek, Michał and Lauritsen, Kristian and Bülter, Andreas and Schneider, Philipp-Immanuel and Zschiedrich, Lin and Burger, Sven and Rodt, Sven and Urbańczyk, Wacław and Sęk, Grzegorz and Reitzenstein, Stephan},
title = {Plug\&Play Fiber-Coupled 73 kHz Single-Photon Source Operating in the Telecom O-Band},
journal = {Advanced Quantum Technologies},
volume = {3},
number = {6},
pages = {2000018},
doi = {https://doi.org/10.1002/qute.202000018},
url = {https://advanced.onlinelibrary.wiley.com/doi/abs/10.1002/qute.202000018},
year = {2020}
}

@article{Yuan2023,
  title = {{GaAs} quantum dots under quasiuniaxial stress: Experiment and theory},
  author = {Yuan, Xueyong and Covre da Silva, Saimon F. and Csontosov\'a, Diana and Huang, Huiying and Schimpf, Christian and Reindl, Marcus and Lu, Junpeng and Ni, Zhenhua and Rastelli, Armando and Klenovsk\'y, Petr},
  journal = {Phys. Rev. B},
  volume = {107},
  issue = {23},
  pages = {235412},
  numpages = {12},
  year = {2023},
  month = {Jun},
  publisher = {American Physical Society},
  doi = {10.1103/PhysRevB.107.235412}
}

@article{VurgaftmanJAP2001,
	author = {I. Vurgaftman and J. R. Meyer and L. R. Ram-Mohan},
	title = {Band parameters for {III–V} compound semiconductors and their alloys},
	journal = {J. Appl. Phys.},
	volume = {89},
	number = {11},
	pages = {5815-5875},
	year = {2001},
	doi = {10.1063/1.1368156}
}

@article{SaidiJAP2010,
	author = {I. Sa{\"{i}}di and S. Ben Radhia and K. Boujdaria},
	title = {Band parameters of {GaAs}, {InAs}, {InP}, and {InSb} in the 40-band $k{\cdot}p$ model},
	journal = {J. Appl. Phys.},
	volume = {107},
	number = {4},
	pages = {043701},
	year = {2010},
	DOI = {10.1063/1.3295900}
}

@article{TseJAP2013,
	author = {G. Tse and J. Pal and U. Monteverde and R. Garg and V. Haxha and M. A. Migliorato and S. Tomi{\'c}},
	title = {Non-linear piezoelectricity in zinc blende {GaAs} and {InAs} semiconductors},
	journal = {J. Appl. Phys.},
	volume = {114},
	number = {7},
	pages = {073515},
	year = {2013},
	DOI = {10.1063/1.4818798}
}

@inbook{AmirtharajBOOK1994,
	author={Paul M. Amirtharaj and David G. Seiler},
	editor={Michael Bass and Eric W.~Van Stryland and David R.~Williams and William L.~Wolfe},
	isbn = {0070479747},
	title={Optical Properties of Semiconductors},
	publisher = {McGraw-Hill Professional},
	booktitle = {Handbook of Optics, Vol. 2: Devices, Measurements, and Properties, Second Edition},
	year = {1994}
}

@article{Gawelczyk2017,
  title = {Exciton lifetime and emission polarization dispersion in strongly in-plane asymmetric nanostructures},
  author = {Gawe\l{}czyk, M. and Syperek, M. and Mary\ifmmode \acute{n}\else \'{n}\fi{}ski, A. and Mrowi\ifmmode \acute{n}\else \'{n}\fi{}ski, P. and Dusanowski, \L{}. and Gawarecki, K. and Misiewicz, J. and Somers, A. and Reithmaier, J. P. and H\"ofling, S. and S\ifmmode \mbox{\k{e}}\else \k{e}\fi{}k, G.},
  journal = {Phys. Rev. B},
  volume = {96},
  issue = {24},
  pages = {245425},
  numpages = {11},
  year = {2017},
  month = {Dec},
  publisher = {American Physical Society},
  doi = {10.1103/PhysRevB.96.245425}
}

@article{Bahder1992,
  title = {Eight-band k\ensuremath{\cdot}p model of strained zinc-blende crystals},
  author = {Bahder, Thomas B.},
  journal = {Phys. Rev. B},
  volume = {41},
  issue = {17},
  pages = {11992--12001},
  numpages = {0},
  year = {1990},
  month = {Jun},
  publisher = {American Physical Society},
  doi = {10.1103/PhysRevB.41.11992},
  url = {https://link.aps.org/doi/10.1103/PhysRevB.41.11992}
}

@article{Mielnik2018,
  title = {Dominant role of the shear strain induced admixture in spin-flip processes in self-assembled quantum dots},
  author = {Mielnik-Pyszczorski, Adam and Gawarecki, Krzysztof and Gawe{\l}czyk, Micha{\l} and Machnikowski, Pawe{\l}},
  journal = {Phys. Rev. B},
  volume = {97},
  issue = {24},
  pages = {245313},
  numpages = {9},
  year = {2018},
  month = {Jun},
  publisher = {American Physical Society},
  doi = {10.1103/PhysRevB.97.245313},
  url = {https://link.aps.org/doi/10.1103/PhysRevB.97.245313}
}

@article{Gawarecki2014,
  title = {Electron states in a double quantum dot with broken axial symmetry},
  author = {Gawarecki, Krzysztof and Machnikowski, Pawe\l{} and Kuhn, Tilmann},
  journal = {Phys. Rev. B},
  volume = {90},
  issue = {8},
  pages = {085437},
  numpages = {8},
  year = {2014},
  month = {Aug},
  publisher = {American Physical Society},
  doi = {10.1103/PhysRevB.90.085437},
  url = {https://link.aps.org/doi/10.1103/PhysRevB.90.085437}
}

@article{Bryant1987,
  title = {Electronic structure of ultrasmall quantum-well boxes},
  author = {Bryant, Garnett W.},
  journal = {Phys. Rev. Lett.},
  volume = {59},
  issue = {10},
  pages = {1140--1143},
  numpages = {0},
  year = {1987},
  month = {Sep},
  publisher = {American Physical Society},
  doi = {10.1103/PhysRevLett.59.1140},
  url = {https://link.aps.org/doi/10.1103/PhysRevLett.59.1140}
}

@article{PhysRevB.65.195313,
  title = {Theory of pure dephasing and the resulting absorption line shape in semiconductor quantum dots},
  author = {Krummheuer, B. and Axt, V. M. and Kuhn, T.},
  journal = {Phys. Rev. B},
  volume = {65},
  issue = {19},
  pages = {195313},
  numpages = {12},
  year = {2002},
  month = {May},
  publisher = {American Physical Society},
  doi = {10.1103/PhysRevB.65.195313},
  url = {https://link.aps.org/doi/10.1103/PhysRevB.65.195313}
}

@article{rickert_high_purcell,
author = {Rickert, Lucas and Vajner, Daniel A. and von Helversen, Martin and Schall, Johannes and Rodt, Sven and Reitzenstein, Stephan and Liu, Hanqing and Li, Shulun and Ni, Haiqiao and Niu, Zhichuan and Heindel, Tobias},
title = {High Purcell Enhancement in Quantum-Dot Hybrid Circular Bragg Grating Cavities for  GHz Clock Rate Generation of Indistinguishable Photons},
journal = {ACS Photonics},
volume = {12},
number = {1},
pages = {464-475},
year = {2025},
doi = {10.1021/acsphotonics.4c01873},
URL = { 
    
        https://doi.org/10.1021/acsphotonics.4c01873
    
    

}

}

@article{williamson_electronic,
  title = {Theoretical interpretation of the experimental electronic structure of lens-shaped self-assembled {InAs/GaAs} quantum dots},
  author = {Williamson, A. J. and Wang, L. W. and Zunger, Alex},
  journal = {Phys. Rev. B},
  volume = {62},
  issue = {19},
  pages = {12963--12977},
  numpages = {0},
  year = {2000},
  month = {Nov},
  publisher = {American Physical Society},
  doi = {10.1103/PhysRevB.62.12963},
  url = {https://link.aps.org/doi/10.1103/PhysRevB.62.12963}
}

@article{Navarez_electronicstructure,
    author = {Narvaez, Gustavo A. and Bester, Gabriel and Zunger, Alex},
    title = {Dependence of the electronic structure of self-assembled {(In,Ga)As/GaAs} quantum dots on height and composition},
    journal = {Journal of Applied Physics},
    volume = {98},
    number = {4},
    pages = {043708},
    year = {2005},
    month = {08},
    issn = {0021-8979},
    doi = {10.1063/1.1980534},
    url = {https://doi.org/10.1063/1.1980534},
}

@article{ramsay_rabi_dampening,
  title = {Damping of Exciton Rabi Rotations by Acoustic Phonons in Optically Excited $\mathrm{InGaAs}/\mathrm{GaAs}$ Quantum Dots},
  author = {Ramsay, A. J. and Gopal, Achanta Venu and Gauger, E. M. and Nazir, A. and Lovett, B. W. and Fox, A. M. and Skolnick, M. S.},
  journal = {Phys. Rev. Lett.},
  volume = {104},
  issue = {1},
  pages = {017402},
  numpages = {4},
  year = {2010},
  month = {Jan},
  publisher = {American Physical Society},
  doi = {10.1103/PhysRevLett.104.017402},
  url = {https://link.aps.org/doi/10.1103/PhysRevLett.104.017402}
}

@article{castelletto2021silicon,
  title={Silicon carbide single-photon sources: challenges and prospects},
  author={Castelletto, Stefania},
  journal={Materials for Quantum Technology},
  volume={1},
  number={2},
  pages={023001},
  year={2021},
  publisher={IOP Publishing},
  doi={10.1088/2633-4356/abe04a}
}

@article{Kaldewey_coherent,
  title = {Coherent and robust high-fidelity generation of a biexciton in a quantum dot by rapid adiabatic passage},
  author = {Kaldewey, Timo and L\"uker, Sebastian and Kuhlmann, Andreas V. and Valentin, Sascha R. and Ludwig, Arne and Wieck, Andreas D. and Reiter, Doris E. and Kuhn, Tilmann and Warburton, Richard J.},
  journal = {Phys. Rev. B},
  volume = {95},
  issue = {16},
  pages = {161302(R)},
  numpages = {5},
  year = {2017},
  month = {Apr},
  publisher = {American Physical Society},
  doi = {10.1103/PhysRevB.95.161302},
  url = {https://link.aps.org/doi/10.1103/PhysRevB.95.161302}
}

@article{Hanschke_reappearence,
  title = {Experimental Measurement of the Reappearance of Rabi Rotations in Semiconductor Quantum Dots},
  author = {Hanschke, Lukas and Bracht, Thomas K. and Sch\"oll, Eva and Bauch, David and Berger, Eva and Kallert, Patricia and Peter, Melina and Garcia, Ailton J. and Covre da Silva, Saimon F. and Manna, Santanu and Rastelli, Armando and Schumacher, Stefan and Reiter, Doris E. and J\"ons, Klaus D.},
  journal = {Phys. Rev. Lett.},
  volume = {135},
  issue = {26},
  pages = {263602},
  numpages = {7},
  year = {2025},
  month = {Dec},
  publisher = {American Physical Society},
  doi = {10.1103/s212-43gs},
  url = {https://link.aps.org/doi/10.1103/s212-43gs}
}

@article{Reiter01012019,
author = {D. E. Reiter and T. Kuhn and V. M. Axt},
title = {Distinctive characteristics of carrier-phonon interactions in optically driven semiconductor quantum dots},
journal = {Advances in Physics: X},
volume = {4},
number = {1},
pages = {1655478},
year = {2019},
publisher = {Taylor \& Francis},
doi = {10.1080/23746149.2019.1655478},


URL = { 
    
        https://doi.org/10.1080/23746149.2019.1655478
}

}

@article{krummheuer2002theory,
  title={Theory of pure dephasing and the resulting absorption line shape in semiconductor quantum dots},
  author={Krummheuer, Birgit and Axt, Vollrath Martin and Kuhn, Tilmann},
  journal={Physical Review B},
  volume={65},
  number={19},
  pages={195313},
  year={2002},
  url={https://journals.aps.org/prb/abstract/10.1103/PhysRevB.65.195313}
}

\end{document}